\documentclass{caosp305}
\usepackage{graphicx}

\usepackage{natbib}
\articleNo{123}
\pubyear{2005}
\volume{35}
\volnumber{3}
\firstpage{1}
\received{May 1, 2005}
\accepted{August 28, 2005}

\def\BibTeX{{\rm B\kern-.05em{\sc i\kern-.025em b}\kern-.08em
             T\kern-.1667em\lower.7ex\hbox{E}\kern-.125emX}}

\begin{document}

%%%%%%%%%%%%%%%%%%%%%%%%%%%%%%%%%%%%%%%%%%%%%%%%%%%%%%%%%%%%%%%%%%%%%%%%%%%%%
%              R U N N I N G   P A G E   H E A D I N G S                     
% Odd page headings (except for the title page) are produced automatically
% and contain the title. If, and only if, the title of your article is too
% long the running head is omitted in the printout; you can make your own
% running title by using the \htitle command, putting the shortened title
% between the curly brackets. \htitle should also be used when the
% subtitle is present: \htitle offers you a way how to include it into the
% headings. If you wish to see how it works simply remove the % sign from
% the beginning of that line.
%
% Unlike the \htitle command, the \hauthor command is compulsory. It is
% used to produce even page headings and contains the names of the authors
% of an article.  All authors must be listed here, if possible. When
% authors' list is too long, you can abbreviate it by using "{\it et
% al.}". Authors' names are given in the form: initial(s) of the author's
% first name and surname. Authors are separated by a "," (comma) sign and
% the last one by "and".
%%%%%%%%%%%%%%%%%%%%%%%%%%%%%%%%%%%%%%%%%%%%%%%%%%%%%%%%%%%%%%%%%%%%%%%%%%%%%
%\htitle{A note to comet ejection process ...}
\hauthor{G.A. Wade \& C. Neiner}
%\hauthor{L.\,Neslu\v{s}an {\it et al.}}

%%%%%%%%%%%%%%%%%%%%%%%%%%%%%%%%%%%%%%%%%%%%%%%%%%%%%%%%%%%%%%%%%%%%%%%%%%%%%
%                       T I T L E                                            
% Capital letters in the title are only used at the beginning of the
% names. Don`t end the title by a "." (dot)
%%%%%%%%%%%%%%%%%%%%%%%%%%%%%%%%%%%%%%%%%%%%%%%%%%%%%%%%%%%%%%%%%%%%%%%%%%%%%
\title{Magnetism of hot stars}

%%%%%%%%%%%%%%%%%%%%%%%%%%%%%%%%%%%%%%%%%%%%%%%%%%%%%%%%%%%%%%%%%%%%%%%%%%%%%
%                       S U B T I T L E                                      
% You can use the subtitle, with the command \subtitle similar to the
% \title command.
%%%%%%%%%%%%%%%%%%%%%%%%%%%%%%%%%%%%%%%%%%%%%%%%%%%%%%%%%%%%%%%%%%%%%%%%%%%%%

%%%%%%%%%%%%%%%%%%%%%%%%%%%%%%%%%%%%%%%%%%%%%%%%%%%%%%%%%%%%%%%%%%%%%%%%%%%%%
%                   A U T H O R  N A M E S                                   
% Authors' names are separated by the \and command and their institutes
% are assigned by the \inst{n} command.
%
% When the name contains "Slovak" letters L,d,t,l with a caron, use an
% a new \softl, etc. command (examples given in the last table of
% this document) to produce typographically correct accented characters.
%%%%%%%%%%%%%%%%%%%%%%%%%%%%%%%%%%%%%%%%%%%%%%%%%%%%%%%%%%%%%%%%%%%%%%%%%%%%%
\author{
        G.A. Wade \inst{1} 
      \and 
        C. Neiner \inst{2}   
       }

%%%%%%%%%%%%%%%%%%%%%%%%%%%%%%%%%%%%%%%%%%%%%%%%%%%%%%%%%%%%%%%%%%%%%%%%%%%%%
%           I N S T I T U T E S'  A D D R E S S E S                          
% The affiliation of authors is generated by the \institute command, the
% \and command being again used to separate individual addresses.
% The following commands may be used for the following three institutes:   
%               \lomnica        for      AsU SAV, Tatranska Lomnica          
%               \blava          for      AsU SAV, Bratislava                 
%               \ondrejov       for      AsU CAV, Ondrejov                   
%
% The given postal address must be complete in order to facilitate our
% editorial work. Moreover, you can add your e-mail address, using the
% \email command.
%%%%%%%%%%%%%%%%%%%%%%%%%%%%%%%%%%%%%%%%%%%%%%%%%%%%%%%%%%%%%%%%%%%%%%%%%%%%%
\institute{
  Department of Physics \& Space Science, Royal Military College of Canada, P.O. Box 17000, Station Forces, Kingston, Ontario, Canada, K7K 7B4
  \and
  LESIA, Observatoire de Paris, PSL Research University, CNRS, Sorbonne Universit\'es, UPMC Univ. Paris 06, Univ. Paris Diderot, Sorbonne Paris Cit\'e, 5 place Jules Janssen, 92195 Meudon, France
          }

%%%%%%%%%%%%%%%%%%%%%%%%%%%%%%%%%%%%%%%%%%%%%%%%%%%%%%%%%%%%%%%%%%%%%%%%%%%%%
%                        D A T E / R E C E I V E D                          
% Date inserted here will be the date when your paper was received The
% format is: month (not abbreviated), day, year.
%%%%%%%%%%%%%%%%%%%%%%%%%%%%%%%%%%%%%%%%%%%%%%%%%%%%%%%%%%%%%%%%%%%%%%%%%%%%%
\date{March 8, 2003}
%\date{March 10, 2003}

%%%%%%%%%%%%%%%%%%%%%%%%%%%%%%%%%%%%%%%%%%%%%%%%%%%%%%%%%%%%%%%%%%%%%%%%%%%%%
%                        M A K E T I T L E
% The beginning part (title, author(s), etc.) of your article must be
% closed by the \maketitle command.
%%%%%%%%%%%%%%%%%%%%%%%%%%%%%%%%%%%%%%%%%%%%%%%%%%%%%%%%%%%%%%%%%%%%%%%%%%%%%
\maketitle

%%%%%%%%%%%%%%%%%%%%%%%%%%%%%%%%%%%%%%%%%%%%%%%%%%%%%%%%%%%%%%%%%%%%%%%%%%%%%
%                        A B S T R A C T,  K E Y W O R D S                   
% Here it is shown how to write an abstract.  Keywords should be placed
% within the "abstract" environment using the command \keywords and they
% should be selected from the thesaurus from Astron.  Astrophys.
% Abstracts. They must be separated from each other by -- (two dashes).
%%%%%%%%%%%%%%%%%%%%%%%%%%%%%%%%%%%%%%%%%%%%%%%%%%%%%%%%%%%%%%%%%%%%%%%%%%%%%
\begin{abstract}
Strong, stable, and organised magnetic fields are present at the surfaces of a small fraction of OBA stars. These "fossil fields'' exhibit uniform characteristics in stars over a tremendous range of stellar mass, age, temperature, and rotation rate. In hot O- and B-type stars, these magnetic fields couple efficiently to the stellar radiatively driven winds, strongly influencing stellar mass loss and rotation. In this article we review the characteristics of the known magnetic hot stars, discuss recent discoveries and insights, and describe recent theoretical progress toward understanding basic field properties and the influence of magnetic fields on hot star evolution.

%The subsurface component of the field may well also introduce coupling of the radiative zone layers, efficiently transporting angular momentum and resulting in a rigidly rotating interior. As a consequence, core overshoot and chemical mixing may be strongly influenced. The evolutionary impact of these complex, diverse and fundamental effects is just beginning to be explored, both theoretically and observationally. 
%
%Magnetic fields are also transformed as stars evolve and their internal structure changes, with fossil fields modified by developing convective flows and their associated dynamos, and with dynamo fields becoming "fossilized'' as once-convective regions become radiative. In recent years, the first clear observational evidence of these processes has been presented, including the decline of surface magnetic field strength as stars expand during their MS evolution, the retention of fossil fields remnants in convective red giant atmospheres, and the relaxation of dynamo fields into fossil structures in the cores of red giants.

\keywords{stars: magnetic fields -- stars: evolution -- stars: winds, outflows -- stars: rotation}
\end{abstract}

%%%%%%%%%%%%%%%%%%%%%%%%%%%%%%%%%%%%%%%%%%%%%%%%%%%%%%%%%%%%%%%%%%%%%%%%%%%%%
%                       S E C T I O N I N G                                  
% Any section starts with the command \section as shown below, with the
% title in Initial Capitals and lowercase only. Do not number the sections
% - let LaTeX do that for you - and do not end them by a "." (dot).
%
% The (sub)section titles are typeset in boldface; so, if working in the
% mathematics mode in (sub)section titles, you must use \boldmath and 
% enclose it into curly brackets, e.g. "{\bolmath $R^{2}$}".
%%%%%%%%%%%%%%%%%%%%%%%%%%%%%%%%%%%%%%%%%%%%%%%%%%%%%%%%%%%%%%%%%%%%%%%%%%%%%
\section{Introduction}

Recent large surveys \citep{2016MNRAS.456....2W,2017MNRAS.465.2432G,2015A&A...582A..45F,2017A&A...599A..66S} have investigated the properties of surface magnetism of stars having radiative envelopes, i.e. stars of spectral types A, B and O ranging in effective temperature from roughly 7,000-50,000~K. The largest of these surveys - the Magnetism in Massive Stars (or MiMeS) project \citep{2016MNRAS.456....2W} - obtained magnetic measurements of over 550 bright stars of spectral types B and O. Based on these data, they reported that organized magnetic fields stronger than a few hundred gauss are present at the surfaces of approximately 7\% of such stars \citep[e.g.][]{2017MNRAS.465.2432G}. Their surface dipolar field components range in strength from less than 100~G to well over 10~kG \citep[e.g.][]{2016A&A...587A.126B,2012MNRAS.425.1278W,2013MNRAS.429..398P,2016PhDT.......390S}. Like their lower-mass A-type cousins, their magnetic fields are not symmetric about the stellar rotation axis, leading to rotational modulation of the measured line-of-sight magnetic field and other photospheric and wind diagnostics, phenomena explained by the Oblique Rotator Model \citep{1950MNRAS.110..395S}.

The essential characteristics of these magnetic fields (their organized topologies, their large range of strengths, their stability on timescales of at least decades, and their lack of correlation with basic stellar parameters such as mass and rotation rate) are identical to those of cooler, lower-mass A-type stars, but stand in strong contrast to those of the dynamo-generated fields of cool stars. On the other hand, they bear a number of similarities to the fields of white dwarfs and neutrons stars. These fields are thought to be "fossils'', i.e. the slowly-evolving remnants of magnetic fields generated or accumulated during an earlier phase of stellar evolution. That the properties of these fields remain essentially unchanged in stars ranging in mass from 1.5 to about 60 times that of the Sun \citep{2014IAUS..302..265W,2015ASPC..494...30W}, notwithstanding the important differences in stellar structure that occur over this enormous range of mass, suggests a basic underlying commonality of the formation or evolution of their magnetic fields. Understanding this remarkable observational fact is a key challenge of modern stellar astrophysics.

\subsection{The physics of fossil magnetism}

%\begin{figure}[b]
%% \vspace*{-2.0 cm}
%\begin{center}
%\begin{minipage}{.45\textwidth}
%\hspace*{-0.5 cm}\includegraphics[width=2.7in]{ostars_hrd_bstars.eps}
%% \vspace*{-1.0 cm}
% \caption{HR diagram of the known magnetic B- and O-type stars. (Courtesy of M. Shultz.)}
% \label{hrd_fig}
% \end{minipage}
% \hspace{0.25cm}
% \begin{minipage}{.45\textwidth}
%\includegraphics[width=2.5in]{ostars_apstars_bd_hist.eps}
% \caption{Distribution of the inferred dipolar magnetic fields strengths of known magnetic B and O stars, compared to a sample of magnetic A stars. (Courtesy of M. Shultz.)}
%   \label{dip_fig}
% \end{minipage}
%\end{center}
%\end{figure}

A major breakthrough of the first decades of the 21st century has been to establish that stochastic initial seed magnetic fields in stellar radiative zones relax naturally to long-lived stable configurations with mixed poloidal/toroidal configurations organized on large scales. Numerical and semi-analytic calculations \citep{2004Natur.431..819B,2010A&A...517A..58D,2010ApJ...724L..34D} have shown that the characteristics of the fields predicted at the stellar surface by these models are in good qualitative agreement with those observed in real magnetic O, B and A-type stars (also see Kochukhov, these proceedings).  In the stellar interior, the toroidal fields are predicted to be very strong, and they may have important consequences for chemical and angular momentum transport throughout the radiative envelope. 

This breakthrough has provided the first quantitative physical model for the interpretation and study of fossil stellar magnetism, with the consequence that there now exists a theoretical context for the interpretation of the magnetic fields observed at the surfaces of main sequence (MS) and evolved A, B and O stars, in white dwarfs \citep{2004Natur.431..819B}, and in neutron stars \citep{2006A&A...450.1097B}. It has even spawned the first theoretical speculations aimed at understanding why only a minority (about 10\%) of MS stars exhibit detectable surface magnetism, and to explain the puzzling "magnetic field desert'' phenomenon \citep{2007A&A...475.1053A,2013MNRAS.428.2789B}.

Notwithstanding this important progress, the origins of the seed fields that give rise to the observed fossil fields remain unknown and poorly constrained \citep{2015sf2a.conf..213N}. Currently, there exist three principal hypotheses: conservation of magnetic flux from the interstellar medium (ISM) during star formation, convective dynamos operating during protostellar or pre-MS phases that could enhance the ISM field \citep[e.g.][]{2014IAUS..302...25H}, and stellar mergers early in the formative history \citep[e.g.][]{2009MNRAS.400L..71F,2014IAUS..302....1L}.

\section{Observed properties of magnetic B- and O-type stars}

To date, about 70 early-type magnetic stars with spectral types earlier than B5 or effective temperatures above 15000 K have been confidently identified. Most are relatively bright stars. \citet{2013MNRAS.429..398P}, \citet{2015ASPC..494...30W}, and \citet{2016PhDT.......390S} provide summaries.

\subsection{Fundamental characteristics}

Fossil magnetic fields are observed without any significant change in their general characteristics in stars ranging in spectral type from mid-F to early O. The earliest magnetic stars have spectral types of about O4, although their spectral classification changes somewhat due to the periodic variability of their spectra \citep{1972AJ.....77..312W,2007MNRAS.381..433H}. These most massive magnetic stars have inferred masses of $40-60~M_\odot$. However, because the field characteristics of magnetic O stars are poorly sampled (only about one dozen are known), this does not represent a confident upper limit and higher-mass magnetic stars could exist.

The HR diagram of the known magnetic hot stars \citep{2016PhDT.......390S} shows that the lion's share of the known magnetic A, B and O stars are MS objects. A small number of pre-MS magnetic B-type stars \citep[see e.g.][]{2013MNRAS.429.1001A} and post-MS magnetic O, B and high-mass A stars \citep[see Martin et al., these proceedings,][]{2017MNRAS.471.1926N,2017arXiv171207403M} are also known. 

\subsection{Spectral properties}

The properties of stellar spectra at visible wavelengths change significantly from the early B stars to the mid-O stars. While some of this change is attributable to the evolving radiation field, changing ionization balance and the importance of non-LTE effects, it is also in large part due to the growing influence of the stellar wind. 

As is discussed later in Sect.~\ref{windfield}, a key impact of a magnetic field at the surface of a hot star is the channeling of its outflowing wind. This phenomenon results in the presence of a large quantity of relatively dense magnetically-confined plasma typically located within a few stellar radii of the stellar surface. The presence of this plasma weakly modifies the optical spectra of magnetic B stars, but can fundamentally alter the spectra of magnetic O stars \citep[see e.g.][]{2013MNRAS.429..398P,2015ASPC..494...30W} by introducing many periodically-variable emission lines. At radio wavelengths, magnetic B and O stars exhibit non-thermal emission, likely due to the gyrosynchrotron mechanism \citep[e.g.][]{2004A&A...418..593T,2017MNRAS.465.2160K}. At shorter wavelengths, X-ray emission diagnoses the hot plasma produced as a consequence of the large-scale shocks caused by channelling of the supersonic wind \citep{2016AdSpR..58..680U}.

In addition to wind channeling effects, the magnetic field may also influence transport processes in the photosphere, producing large-scale chemical abundance inhomogeneities \citep[e.g.][]{2015MNRAS.451.2015O}. This chemical spots, which are prevalent for magnetic stars cooler than about 25,000~K, are responsible for periodic modulation of the profiles of many photospheric spectral lines.

\subsection{Magnetic field strengths and topologies}

\citet{2016PhDT.......390S} (see also see Shultz et al., these proceedings) performed a homogeneous study of the physical, rotational, magnetic and magnetospheric properties of the known magnetic B-type stars. The properties of the known magnetic O stars are summarized by \citet{2015ASPC..494...30W}. Apart from a small number of well-known exceptions (e.g. the B stars $\tau$~Sco and HD 37776), the magnetic field toplogies of hot stars appear to be qualitatively identical to those of their lower-mass Ap star cousins: they contain most of their magnetic energy in the dipole mode, their dipole axes are inclined significantly relative to their rotation axes, and their magnetic configurations are stable on long timescales (years to decades, corresponding to hundreds or even thousands of stellar rotations). The distributions of surface dipole magnetic field strengths are very similar for O, B and A stars: they range from a few times $10^1$ to a few times $10^4$ G, and peak at a few times $10^3$~G. 

A key characteristics of the distribution of magnetic fields strengths of Ap stars is the so-called "magnetic desert'', characterized by a "critical dipole field strength'' of about 300~G below which essentially no magnetic stars are detected \citep{2007A&A...475.1053A}. The characteristics of this 'desert' for more massive stars remains to be established, as several clear examples of stars with field strengths below 300~G have already been discovered \citep[e.g.][]{2016A&A...587A.126B}. 

While the magnetic properties of the hot pre-MS magnetic stars are similar to those of the MS population, the evolved post-MS magnetic stars generally have much weaker fields \citep{2017MNRAS.471.1926N}, as would be expected from magnetic flux conservation \citep[e.g.][]{2017IAUS..329..250K}. Those evolved stars thus often appear in the magnetic desert, and explain, at least to some extent, the hot stars that reside there.

\subsection{Rotation}

Magnetic hot stars generally exhibit rotation rates below those of "normal'', non-magnetic stars of similar spectral types. The magnetic field impacts the stellar rotation by coupling to the outflowing wind, enhancing angular momentum loss and rapidly braking the star. Nevertheless, several examples of magnetic hot stars exhibiting rapid rotation have been discovered \citep[e.g.][]{2010MNRAS.405L..51O,2012MNRAS.419.1610G}.

Classical Be stars exhibit the most rapid rotation of any class of B-type stars. As of today, no magnetic field has been directly detected in any
classical Be star, notwithstanding extensive searches \citep{2016MNRAS.456....2W,2016ASPC..506..207W}. However,
indirect evidence of the presence of a magnetic field, i.e. rotationally modulated observables, seems to be have been found in the classical Be star
$\omega$\,Ori \citep{2003A&A...409..275N,2012MNRAS.426.2738N}.

\subsection{Binarity}

Magnetic fields in binary systems may be strongly affected by, and may also strongly affect, the transfer of energy, mass and angular momentum between the components.

Although binary systems containing hot stars are extremely common, binaries with orbital periods shorter than $\sim 60$~days
containing magnetic hot stars are very rare. At the conclusion of the MiMeS survey,
less than one dozen SB2 systems containing magnetic A, B or O stars were known.
Of these, less than one-half contained a hot primary star. The Binarity and
Magnetic Interactions in various classes of Stars \citep[BinaMIcS, e.g.][]{2015sf2a.conf..213N} project has studied
this issue further in short-period spectroscpic binaries and indeed confirms
that the occurence of magnetic fields in hot stars in those binaries is lower
than in single hot stars. A possible explanation for this
dearth of magnetic fields in hot binary systems may lie in stellar formation
processes. For example, \citet{2011ApJ...742L...9C} showed in their simulations
that fragmentation of dense stellar cores is inhibited when the medium is
magnetic. This could make binary system more difficult to form in the presence
of a seed field. 

Similar discrepancies between incidence rates of strong magnetic fields in cataclysmic variables (CV) vs. single white dwarfs led \citet{2008MNRAS.387..897T} to conclude that the formation of white dwarf magnetic fields was initimately tied to the physics of CV formation, in particular the mass transfer and merger processes. Others \citep[e.g.][]{2014IAUS..302....1L} have speculated about a similar connection between binarity and the origin of the fossil magnetic fields of non-degenerate hot stars. The binarity of magnetic hot stars is discussed in greater detail by Naz\'e et al. (these proceedings).

\section{Stellar wind-magnetic field interactions}\label{windfield}

Magnetic stars of spectral types A to moderately early B (often called Ap and Bp stars) exhibit strong, characteristic spectral line strength anomalies and variability. These phenomena - diagnostic of complex, large-scale distributions of abundance enhancement and depletion of various chemical elements - result from the interaction of the magnetic field with photospheric atoms diffusing under the competitive effects of gravity and radiative levitation \citep[e.g.][]{2017MNRAS.468.1023A}. These structures are able to form and subsist because, at the effective temperatures of these stars, the radiative and gravitational forces on some ions are of similar orders of magnitude. At earlier spectral types, the strong growth of the UV radiation field leads to radiative accelerations that rapidly overwhelm gravity, leading to the appearance of radiatively driven stellar winds. (The disappearance of evidence of systematic photospheric chemical peculiarities and abundance structures at about the temperature at which winds become significant is thereby naturally explained.)

Systematic MHD studies of the interaction of these outflowing winds with dipolar magnetic fields have been carried out during the past 15 years \citep[e.g.][]{2002ApJ...576..413U,2008MNRAS.385...97U,2009MNRAS.392.1022U,2013MNRAS.429..398P}). A basic conclusion of these investigations is that two physical quantities are capable of describing the general behaviour of the wind of a hot star under the influence of a magnetic field and stellar rotation (\cite{2013MNRAS.429..398P}): the {\em wind magnetic confinement parameter} \citep{2002ApJ...576..413U}, which determines the {\em Alfv\'en radius} $R_{\rm A}$, and the {\em rotation parameter} \citep{2008MNRAS.385...97U}, which determined the Kepler co-rotation radius $R_{\rm K}$.  
 
In the case of a rapidly-rotating star, the Kepler radius is located relatively close to the stellar surface, and for sufficiently strong magnetic fields is located inside the Alfv\'en radius. In this scenario, plasma in the region between $R_{\rm K}$ and $R_{\rm A}$ is forced (by the magnetic field) to orbit at greater than the local Keplerian speed, and hence experiences an unbalanced (outward) net  force. In such a "centrifugal magnetosphere" (CM), wind plasma is trapped in this region by the combined effects of magnetic field and rotation. In the case of a slowly-rotating star, the Kepler radius is located far from the stellar surface. The net gravitational + centrifugal force is always directed toward the star. In such a "dynamical magnetosphere" (DM) scenario, plasma driven up the field lines ultimately cools and falls back to the stellar surface. The material in the DM is thus frequently renewed (i.e. on the dynamical timescale). 

Centrifugal and dynamical magnetospheres can be observationally distinguished in several ways. First, broad emission features, often found at high velocities, are observed in optical spectra of stars hosting a CM, while stars with DMs show narrower emission if any \citep[see, e.g.][]{2015IAUS..305...53G}. DM emission is mostly observed for O stars, since the winds of B stars are too weak to feed the magnetosphere at a sufficient rate to produce significant emission.  Rotationally-modulated variability of the magnetospheric emission can be used to reconstruct the circumstellar plasma distribution, especially for stars with CMs \citep[e.g.][]{2013msao.confE..69G}. The magnetospheres of hot stars can be comparatively classified using the {\em rotation-confinement diagram} \citep{2013MNRAS.429..398P}.

% FIGURES ILLUSTRATING MDOT QUENCHING AND MAGNETIC BRAKING

\section{The influence of magnetic fields on stellar evolution}

Current 2D and 3D MHD models can effectively compute the short-term evolution of the wind under such conditions, and have provided a sound theoretical basis for understanding the general observational behaviour of hot magnetic stars. In particular, they demonstrate that the surface field interaction with the wind results in two effects that are predicted to significantly influence the evolution of hot stars. {\em Mass-loss quenching} \citep[e.g.][]{2002ApJ...576..413U} refers to the net reduction in the mass loss rate of a star through the top of its magnetosphere as a consequence of magnetic wind trapping. {\em Magnetic braking} \citep[e.g.][]{2009MNRAS.392.1022U} refers to the enhanced loss of angular momentum through the wind resulting principally from Maxwell stresses imparted by the magnetic field. Recent efforts have sought to incorporate the effects of these two important phenomena into models of stellar evolution. \citet{2011A&A...525L..11M} performed first calculations that showed the potential evolutionary impact of magnetic braking.  Keszthelyi et al. (these proceedings) have developed more complete and realistic models, and demonstrate that mass-loss quenching also yields an important evolutionary impact. \citet{2017MNRAS.466.1052P} and \citet{2017A&A...599L...5G} have recently exploited these new modeling capabilities to provide potential explanations of the appearance of high-mass stellar black holes and pair-instability supernovae in high ($\sim$Galactic) metallicity environments. Moreover, recent identification of candidate magnetic stars in the Magellanic Clouds \citep[e.g.][]{2015A&A...577A.107N,2015AJ....150...99W} provides the potential to explore the properties and evolution of magnetic stars in environments very different from those occurring in the Milky Way.

In addition to these predicted (and observed) large scale poloidal surface fields, models of relaxed fossil fields predict very strong, predominantly toroidal fields throughout the stellar radiative zone (see Augustson et al., these proceedings). The extent to which these fields are modified or modify internal circulation currents and differential rotation, including coupling of the rotation of the core to the envelope, is of great current interest \citep{2014IAUS..302....1L} and is beginning to become amenable to observational verification \citep{2012MNRAS.427..483B}. 
 
%
%\begin{figure}[b]
%% \vspace*{-2.0 cm}
%\begin{center}
% \includegraphics[width=2.8in]{fig7.eps}\includegraphics[width=2.82in]{Metal.pdf} 
%% \vspace*{-1.0 cm}
% \caption{The surface wind-magnetic field interaction. {\em Left -}\ From \citet{2009MNRAS.392.1022U}. {\em Right -}\ From \citet{2017MNRAS.466.1052P}.}
%   \label{fig1}
%\end{center}
%\end{figure}

\section{Transformation of fields on evolutionary timescales}

As magnetic fields influence stellar evolution, so are magnetic fields expected to transform in response to changes in the structure of the stars in which they are embedded. The evolution of surface magnetic fields of MS stars has been investigated by e.g. \citet{2007A&A...470..685L}, and more recently by \citet{2016A&A...592A..84F}. These studies suggest that magnetic flux conservation - a basic assumption in models attempting to connect fossil magnetism through different phases of stellar evolution - is poorly supported by observations of main sequence stars.

Extensive surveys of magnetic fields in cool giants and supergiants (the evolutionary descendants of hot MS stars) show that these stars exhibit magnetic fields powered by dynamos \citep[e.g.][]{2010MNRAS.408.2290G,2015A&A...574A..90A}. Although a small population of red giants show evidence of surviving fossil fields from the MS \citep[and the unique interactions between the post-MS dynamo and the pre-existing fossil field; e.g.][]{2008A&A...491..499A}, the growth of the deep convective envelope generally appears to erase evidence of their earlier magnetic characteristics. However, because hot OBA supergiants retain the radiative envelopes they had on the MS, these stars provide a capability to directly extend the studies of MS objects to more advanced evolutionary phases and a much greater range of stellar structural changes \citep[e.g.][]{2015A&A...582A.110B,2017MNRAS.471.1926N}. The Large Impact of magnetic Fields on the Evolution of hot stars (LIFE) project aims at detecting and characterizing magnetic evolved hot stars at various phases of the post-MS evolution. This project and magnetic fields of evolved stars are discussed in more detail by Martin et al. (these proceedings).

%- Augustson core dynamos - this conference
%- Auriere results for RGs
%- Stello/Fuller results for RG cores

\bibliographystyle{caosp}
\bibliography{brno1}

%%%%%%%%%%%%%%%%%%%%%%%%%%%%%%%%%%%%%%%%%%%%%%%%%%%%%%%%%%%%%%%%%%%%%%%%%%%%%
%                       H A P P Y E N D                                      
% Your LaTeX source text must be ended by the line:                          
%%%%%%%%%%%%%%%%%%%%%%%%%%%%%%%%%%%%%%%%%%%%%%%%%%%%%%%%%%%%%%%%%%%%%%%%%%%%%
\end{document}